\documentclass[aps,prb,preprint,superscriptaddress,showpacs,floatfix]{revtex4}

\usepackage{amsmath}
\usepackage{graphicx}

\newcommand{\kvec}{\textbf{k}}
\newcommand{\xvec}{\textbf{x}}
\newcommand{\Jvecrt}{\textbf{J}(\xvec,t)}
\newcommand{\Bfieldrt}{\textbf{B}(\xvec,t)}
\newcommand{\Hfieldrt}{\textbf{H}(\xvec,t)}
\newcommand{\Efieldrt}{\textbf{E}(\xvec,t)}

\newcommand{\Hfieldr}{\textbf{H}(\xvec)}
\newcommand{\Efieldr}{\textbf{E}(\xvec)}

\newcommand{\JHDvecr}{\textbf{J}_{\text{HD}}(\xvec)}
\newcommand{\JLjvecr}{\textbf{J}_{\text{L} \textit{j}}(\xvec)}

\newcommand{\partialt}{\frac{\partial}{\partial t}}
\newcommand{\Dfieldrt}{\textbf{D}(\xvec,t)}
\newcommand{\curlF}[1]{\nabla \times #1}

\begin{document}

\title{Calculating Nonlocal Optical Properties of Structures with Arbitrary Shape}

\author{Jeffrey M. McMahon} \email[]{jeffrey-mcmahon@northwestern.edu}
\affiliation{Department of Chemistry, Northwestern University, Evanston, IL 60208}
\affiliation{Center for Nanoscale Materials, Argonne National Laboratory, Argonne, IL 60439}

\author{Stephen K. Gray}
\affiliation{Center for Nanoscale Materials, Argonne National Laboratory, Argonne, IL 60439}

\author{George C. Schatz}
\affiliation{Department of Chemistry, Northwestern University, Evanston, IL 60208}

\date{\today}

\begin{abstract}
In a recent Letter [\textit{Phys. Rev. Lett.} \textbf{103}, 097403 (2009)], we outlined a computational method to calculate the optical properties of structures with a spatially nonlocal dielectric function. In this Article, we detail the full method, and verify it against analytical results for cylindrical nanowires. Then, as examples of our method, we calculate the optical properties of Au nanostructures in one, two, and three dimensions. We first calculate the transmission, reflection, and absorption spectra of thin films. Because of their simplicity, these systems demonstrate clearly the longitudinal (or volume) plasmons characteristic of nonlocal effects, which result in anomalous absorption and plasmon blueshifting. We then study the optical properties of spherical nanoparticles, which also exhibit such nonlocal effects. Finally, we compare the maximum and average electric field enhancements around nanowires of various shapes to local theory predictions. We demonstrate that when nonlocal effects are included, significant decreases in such properties can occur. 
\end{abstract}


\maketitle

\section{Introduction}
\label{sec:intro}

Interest in the optical properties of metallic nanostructures has been steadily increasing as experimental techniques for their fabrication and investigation have become more sophisticated \cite{VanDuyne_LSPR-Overview_2007}. One of the main driving forces of this is their potential utility in sensing, photonic, and optoelectronics applications \cite{VanDuyne_LSPR-Overview_2007, plasmonic-sensors_SKG_2008, Ozbay_nanoelectronics_2006}. However, there can also be interesting fundamental issues to consider, particularly as very small length scales are approached (approximately less than 10 nm). In this limit, quantum mechanical effects can lead to unusual optical properties relative to predictions based on classical electrodynamics applied with bulk, local dielectric values for the metal \cite{Opt-Prop_Clusters_Vollmer}. In isolated spherical nanoparticles, for example, localized surface plasmon resonances (LSPRs) are found to be blueshifted relative to Mie theory predictions \cite{Novotny_2006}, and in thin metal films, anomalous absorption is observed \cite{NL-exp_Fitton_1971,NL-exp_Lindau_1970}.

Roughly speaking, when light interacts with a structure of size $d$ (e.g., a nanoparticle size or junction gap distance), wavevector components $\kvec$, which are related to the momentum $\textbf{p}$ by $\textbf{p} = \hbar \kvec$, where $\hbar$ is Planck's constant, are generated with magnitude $k = 2 \pi / d$. This, in turn, imparts an energy of $E = (\hbar \kvec)^2 / 2m_e$, where $m_e$ is the mass of an electron, to (relatively) free electrons in the metal. For small $d$, these energies can correspond to the optical range (1 -- 6 eV). This analysis suggests that such effects should come into play for $d$ less than approximately 2 nm. In metals, however, somewhat larger $d$ values also exhibit these effects, because electrons in motion at the Fermi velocity can be excited by the same energy with a smaller momentum increase, due to dispersion effects. 

A full quantum mechanical treatment of such structures would of course be best, but this is not practical for these sizes. However, it is possible to incorporate some ``quantum effects'' within classical electrodynamics via use of a different dielectric model than that for the bulk metal. At least four such effects can be addressed in this way: electron scattering, electron spill-out, quantum size-effects, and spatial nonlocality of the material polarization. The additional losses due to increased electron scattering at the metal surface can be described by a size-dependent damping term \cite{Schatz_size-dep-damping}, which will effectively broaden spectral peaks \cite{McMahon_NL_nanoshells}, as we consider below. The electron spill-out from the metal into the medium, due to the electron density varying smoothly, can partially be accounted for by a dielectric layer model. The effect of this is varied, and depends on a number of details, including the surface chemistry of the structure \cite{McMahon_Ag-blueshift} and its local dielectric environment.
Quantum size effects due to discrete electronic energy levels can lead to a size and shape-dependent conductivity. At least for metal films \cite{metal-films_quantum-effects_Ashcroft_PRB_1988}, this quantity is reduced relative to the bulk, and exhibits peaks for certain film thicknesses. Such effects can be incorporated directly into classical calculations for some simple systems, based on rigorous theory. Although, because this effect and electron spill-out are both highly dependent on system specifics, incorporating them into a general framework is not straightforward. Therefore, they will not be considered in this work. The fourth quantum effect, and the one that is of main interest herein, is the need of a dielectric model which considers that the material polarization at one point in space depends not only on the local electric field, but also that in its neighborhood \cite{Wolf_1974}. 

In classical electrodynamics, materials are described through a dielectric function $\varepsilon$ that relates the electric displacement field $\textbf{D}$ (proportional to both the incident field and material polarization) to the electric field $\textbf{E}$ at a given frequency of light $\omega$. This relationship is usually assumed to be  local in space (i.e., the polarizability of the material at a point $\textbf{x}$ only depends on $\textbf{E}$ at $\textbf{x}$). However, in the more general case
\begin{equation}
  \label{eq:E-to-D}
  \textbf{D}(\xvec, \omega) = \varepsilon_0 \int d \xvec' \varepsilon (\xvec, \xvec', \omega) \textbf{E}(\xvec', \omega) ~~ ,
\end{equation}
where $\varepsilon (\textbf{x}, \textbf{x}', \omega)$ is a spatially-dependent (nonlocal) and frequency-dispersive relative dielectric function. In a homogeneous environment (an approximation which we make for the finite, arbitrarily shaped structures considered herein), $\varepsilon (\textbf{x}, \textbf{x}', \omega)$ only spatially depends on $| \xvec - \xvec' |$. Therefore, in $\kvec$-space, Eq. (\ref{eq:E-to-D}) is more simply expressed as 

\begin{equation}
  \label{eq:E-to-D-k}
  \textbf{D}(\kvec, \omega) = \varepsilon_0  \varepsilon (\kvec, \omega) \textbf{E}(\kvec, \omega) ~~ .
\end{equation}

Since the first formulation of nonlocal electromagnetics \cite{Wolf_1974}, applications of $\kvec$-dependent dielectric functions have remained limited to simple systems, such as spherical structures \cite{Fuchs_1981, Leung_nonlocal_nanoshells} or aggregates thereof \cite{Abajo_2008, Stefanou_2008, Wannemacher_2001, Yannopapas_2008}, and planar surfaces \cite{Fuchs_metal-films_1969}. Nonetheless, this $\kvec$-dependence has been found experimentally \cite{NL-exp_Fitton_1971,NL-exp_Lindau_1970} and proven theoretically \cite{Fuchs_metal-films_1969, Fuchs_1981} to have important consequences. For example, such dependence is responsible for the aforementioned anomalous absorption and LSPR blueshifting. 

In a recent Letter \cite{McMahon_NLDiel}, we outlined a powerful, yet simple method by which the optical properties of arbitrarily shaped structures with a nonlocal dielectric function can easily be calculated. This was done by deriving an equation of motion for the current associated with the hydrodynamic Drude model \cite{Boardman_1982}, which we solved within the framework of the finite-difference time-domain (FDTD) method \cite{Taflove_FDTD}. The advantage of this approach is that it can describe the dynamical optical response of structures that are too large to treat using quantum mechanics, yet small enough such that the application of local continuum electrodynamics becomes questionable.
In this Article, we expand on that work and detail the full method. We first verify our results against analytical ones for the cylindrical Au nanowires that we considered in our previous Letter \cite{McMahon_NLDiel}. Then, as new examples, we calculate the optical properties of one, two, and three dimensional Au nanostructures. 

\section{Theoretical Approach}
\label{sec:method}

\subsection{Formulation of the Method}
\label{sec:method:formulation}

The interaction of light with matter in the classical continuum limit (i.e., many hundreds of atoms or more) is described by Maxwell's equations,
\begin{equation}
  \label{eq:MaxwellAmperert}
  \partialt \Dfieldrt + \Jvecrt = \curlF{\Hfieldrt}
\end{equation}
\begin{equation}
  \label{eq:FaradayLawrt}
  \partialt \Bfieldrt = - \curlF{\Efieldrt}
\end{equation}
\begin{equation}
  \label{eq:Gaussrt}
  \nabla \cdot \Dfieldrt = \rho
\end{equation}
\begin{equation}
  \label{eq:GaussMagrt}
  \nabla \cdot \Bfieldrt = 0
\end{equation}
where $\Hfieldrt$ and $\Bfieldrt$ are the auxiliary magnetic field and \emph{the} magnetic field, respectively, while $\Jvecrt$ and $\rho$ are external current and charge densities. Except for the most simple systems, such as spheres or metal films, analytical solutions or simplifying approximations to Eqs.\ (\ref{eq:MaxwellAmperert}) -- (\ref{eq:GaussMagrt}) do not exist. Therefore, computational methods are often used to solve them, one of the most popular being FDTD \cite{Taflove_FDTD}. For dynamical fields, Eqs.\ (\ref{eq:MaxwellAmperert}) and (\ref{eq:FaradayLawrt}) are explicitly solved, while Eqs.\ (\ref{eq:Gaussrt}) and (\ref{eq:GaussMagrt}) are considered initial conditions that should remain satisfied for all time.

However, before Maxwell's equations can be solved, an explicit form for $\varepsilon (\kvec, \omega)$ in the constitutive relationship between $\textbf{D}(\kvec, \omega)$ and $\textbf{E}(\kvec, \omega)$, Eq. (\ref{eq:E-to-D-k}), must be specified. Note that Eqs.\ (\ref{eq:MaxwellAmperert}) -- (\ref{eq:GaussMagrt}) are in terms of $\xvec$ and $t$, but material properties are often dependent on $\kvec$ and $\omega$, which are related to the former via Fourier transform. Also note that we assume that there are no magnetic materials present, and thus the magnetic field constitutive relationship is $\textbf{B}(\xvec, \omega) = \mu_0 \textbf{H}(\xvec, \omega)$, where $\mu_0$ is the vacuum permeability. Returning to the current discussion, the dielectric function of a metal like Au is well-described in the classical continuum limit by three separate components,
\begin{equation}
  \label{eq:electric_permittivity}
  \varepsilon(\textbf{k}, \omega) = \varepsilon_{\infty} + \varepsilon_\text{inter}(\omega) + \varepsilon_\text{intra}(\textbf{k}, \omega) ~~ ,
\end{equation}
the value as $\omega\rightarrow\infty$, $\varepsilon_{\infty}$, a contribution from \textit{d}-band to \textit{sp}-band (conduction band) interband electron transitions, $\varepsilon_\text{inter}(\omega)$, and a contribution due to \textit{sp}-band electron excitations, $\varepsilon_\text{intra}(\textbf{k}, \omega)$. The notation in Eq.\ (\ref{eq:electric_permittivity}) highlights the $\textbf{k}$ and $\omega$ dependencies.

$\varepsilon_\text{inter}(\omega)$ can be physically described using a multipole Lorentz oscillator model \cite{Light-Scattering_BH},
\begin{equation}
  \label{eq:epsilon_Lorentz}
  \varepsilon_\text{inter}(\omega) = \sum_j \frac{\Delta \varepsilon_{\text{L}j} \omega_{\text{L}j}^2}{\omega_{\text{L}j}^2 - \omega \left( \omega + i 2 \delta_{\text{L}j} \right)}
\end{equation}
where $j$ is an index labeling the individual \textit{d}-band to \textit{sp}-band electron transitions occurring at $\omega_{\text{L}j}$, $\Delta \varepsilon_{\text{L}j}$ is the shift in relative permittivity at the transition, and $\delta_{\text{L}j}$ is the electron dephasing rate. Because there are two interband transitions in Au at optical frequencies (near 3 and 4 eV \cite{OptConst_NobleMetals_JC}), we take $j = 2$ in this work. 

$\varepsilon_{\text{intra}}(\textbf{k}, \omega)$ is responsible for both the plasmonic optical response of metals and nonlocal effects. Both of these can be described by the hydrodynamic Drude model \cite{Boardman_1982}, which reduces to the local Drude expression for electron motion if $\textbf{k} \rightarrow 0$ \cite{Light-Scattering_BH},
\begin{equation}
  \label{eq:hydrodynamic-Drude}
  \varepsilon_\text{intra}(\textbf{k}, \omega) = -\frac{\omega_{\text{D}}^{2}}{\omega(\omega+i\gamma)-\beta^2\textbf{k}^2}
\end{equation}
where $\omega_{\text{D}}$ is the plasma frequency, $\gamma$ is the collision frequency, and for a free electron gas (i.e., one with only kinetic energy) $\beta^2 = C v_\text{F}^2 / D$, where $v_\text{F}$ is the Fermi velocity ($1.39 \cdot 10^6$ m/s for Au), $D$ is the dimension of the system, and $C = 1$ at low frequencies and $3D/(D + 2)$ at high frequencies \cite{Fetter_1973} \footnote{In our previous Letter \cite{McMahon_NLDiel}, the low-frequency 2D value of $\beta^2$ was used.}.
We note, in passing, that other analytical forms for $\varepsilon_{\text{intra}}(\textbf{k}, \omega)$ could also be used with the following approach, such as those inferred from representative quantum mechanical electronic structure calculations \cite{QM_nonlocal-diel_Rubio_2008}.

Inserting Eqs.\ (\ref{eq:E-to-D-k}) and (\ref{eq:electric_permittivity}) [using Eqs.\ (\ref{eq:epsilon_Lorentz}) and (\ref{eq:hydrodynamic-Drude})] into the Maxwell--Amp\`{e}re law in $\kvec$-space for a time-harmonic field, $-i \omega \textbf{D}(\textbf{k}, \omega) = i\textbf{k} \times \textbf{H}(\textbf{k}, \omega)$, leads to
\begin{equation}
  \label{eq:HD2L_AmpereLaw_komega}
  -i \omega \varepsilon_0 \varepsilon_{\infty} \textbf{E}(\textbf{k}, \omega) + \sum_j \textbf{J}_{\text{L}j}(\textbf{k}, \omega) + \textbf{J}_{\text{HD}}(\textbf{k}, \omega) = i\textbf{k} \times \textbf{H}(\textbf{k}, \omega) ~~ ,
\end{equation}
where the $\textbf{J}_{\text{L}j}(\textbf{k}, \omega)$ are polarization currents associated with Eq.\ (\ref{eq:epsilon_Lorentz}),
\begin{equation}
  \label{eq:current_Lorentz_phasor}
  \textbf{J}_{\text{L}j}(\textbf{k}, \omega) = - i \omega \varepsilon_0 \frac{\Delta \varepsilon_{\text{L} j} \omega_{\text{L}j}^2}{\omega_{\text{L}j}^2 - \omega \left( \omega + i 2 \delta_{\text{L}j} \right) } \textbf{E}(\textbf{k}, \omega) ~~ ,
\end{equation}
and $\textbf{J}_{\text{HD}}(\textbf{k}, \omega)$ is a nonlocal polarization current associated with Eq.\ (\ref{eq:hydrodynamic-Drude}),
\begin{equation}
  \label{eq:current_HDrude_phasor}
  \textbf{J}_{\text{HD}}(\textbf{k}, \omega) = i \omega \varepsilon_0 \frac{\omega_{\text{D}}^{2}}{\omega(\omega+i\gamma)-\beta^{2}\textbf{k}^{2}}\textbf{E}(\textbf{k}, \omega) ~~ .
\end{equation}
Equations of motion (i.e., partial differential equations in terms of $\xvec$ and $t$) for the currents in Eqs.\ (\ref{eq:current_Lorentz_phasor}) and (\ref{eq:current_HDrude_phasor}) can be obtained by multiplying through each by the appropriate denominator and inverse Fourier transforming ($i\textbf{k}\rightarrow\nabla$ and $-i\omega\rightarrow\partial / \partial t$),
\begin{equation}
  \label{eq:current_Lorentz_TD-ADE}
  \frac{\partial^{2}}{\partial t^{2}}\textbf{J}_{\text{L}j}(\xvec, t) + 2 \delta_{\text{L}j} \frac{\partial}{\partial t}\textbf{J}_{\text{L}j}(\xvec, t) + \omega_{\text{L}j}^2 \textbf{J}_{\text{L}j}(\xvec, t) = \varepsilon_{0} \Delta \varepsilon_{\text{L}j} \omega_{\text{L}j}^{2} \frac{\partial}{\partial t}\textbf{E}(\xvec, t)
\end{equation}
\begin{equation}
  \label{eq:current_HD_TD-ADE}
  \frac{\partial^{2}}{\partial t^{2}}\textbf{J}_{\text{HD}}(\xvec, t) + \gamma\frac{\partial}{\partial t}\textbf{J}_{\text{HD}}(\xvec, t) - \beta^{2}\nabla^{2}\textbf{J}_{\text{HD}}(\xvec, t) = \varepsilon_{0}\omega_{\text{D}}^{2}\frac{\partial}{\partial t}\textbf{E}(\xvec, t) ~~ .
\end{equation}
Because of the spatial derivatives in Eq.\ (\ref{eq:current_HD_TD-ADE}), the equation of motion for the hydrodynamic Drude model is second-order, unlike the normal Drude model that is first-order \cite{Taflove_FDTD}.

Equations (\ref{eq:current_Lorentz_TD-ADE}) and (\ref{eq:current_HD_TD-ADE}) can be solved self-consistently with Eq.\ (\ref{eq:FaradayLawrt}) and the inverse Fourier-transformed form of Eq.\ (\ref{eq:HD2L_AmpereLaw_komega}),
\begin{equation}
  \label{eq:HD2L_AmpereLaw_rt}
  \varepsilon_0 \varepsilon_{\infty} \partialt \Efieldrt + \sum_j \textbf{J}_{\text{L}j}(t) + \textbf{J}_{\text{HD}}(\xvec, t) = \curlF{\Hfieldrt} ~~ ,
\end{equation}
along with the requirement that Eqs.\ (\ref{eq:Gaussrt}) and (\ref{eq:GaussMagrt}) are, and remain satisfied. Our implementation of Eqs.\ (\ref{eq:FaradayLawrt}) and (\ref{eq:current_Lorentz_TD-ADE}) -- (\ref{eq:HD2L_AmpereLaw_rt}) using standard finite-difference techniques is outlined in the Appendix.

%
%
%
\subsection{Au Dielectric Function}
\label{Section:Diel_Model}

To model Au nanostructures, and use the approach outlined in Subsection \ref{sec:method:formulation}, Eq.\ (\ref{eq:electric_permittivity}) must first be fit to the experimentally determined dielectric data of bulk Au \cite{OptConst_NobleMetals_JC}. This is done in the limit of $\kvec \rightarrow \textbf{0}$ in Eq.\ (\ref{eq:hydrodynamic-Drude}), which is valid for large structures, such as those used to obtain the experimental data. To make sure that the separate terms in Eq.\ (\ref{eq:electric_permittivity}) accurately capture the physics of the problem, it is necessary to fit Eqs.\ (\ref{eq:epsilon_Lorentz}) and (\ref{eq:hydrodynamic-Drude}) over the appropriate energy ranges separately. Using simulated annealing, we first fit Eq.\ (\ref{eq:hydrodynamic-Drude}) (also incorporating $\varepsilon_{\infty}$) over the range 1.0 -- 1.8 eV, where $\varepsilon(\textbf{0}, \omega)$ is dominated by \textit{sp}-band electron motion. Then, keeping the parameters in Eq.\ (\ref{eq:hydrodynamic-Drude}) constant (but not $\varepsilon_{\infty}$), the entire dielectric function in Eq.\ (\ref{eq:electric_permittivity}) was fit over the full range of interest, 1.0 -- 6.0 eV. The resulting parameters were: $\varepsilon_{\infty} = 3.559$, $\omega_{\text{D}} = 8.812$ eV, $\gamma = 0.0752$ eV, $\Delta \varepsilon_{\text{L}1} = 2.912$, $\omega_{\text{L}1} = 4.693$ eV, $\delta_{\text{L}1} = 1.541$ eV, $\Delta \varepsilon_{\text{L}2} = 1.272$, $\omega_{\text{L}2} = 3.112$ eV, and $\delta_{\text{L}2} = 0.525$ eV. 

\begin{figure}
  \includegraphics[scale = 0.45, bb=0 0 656 882]{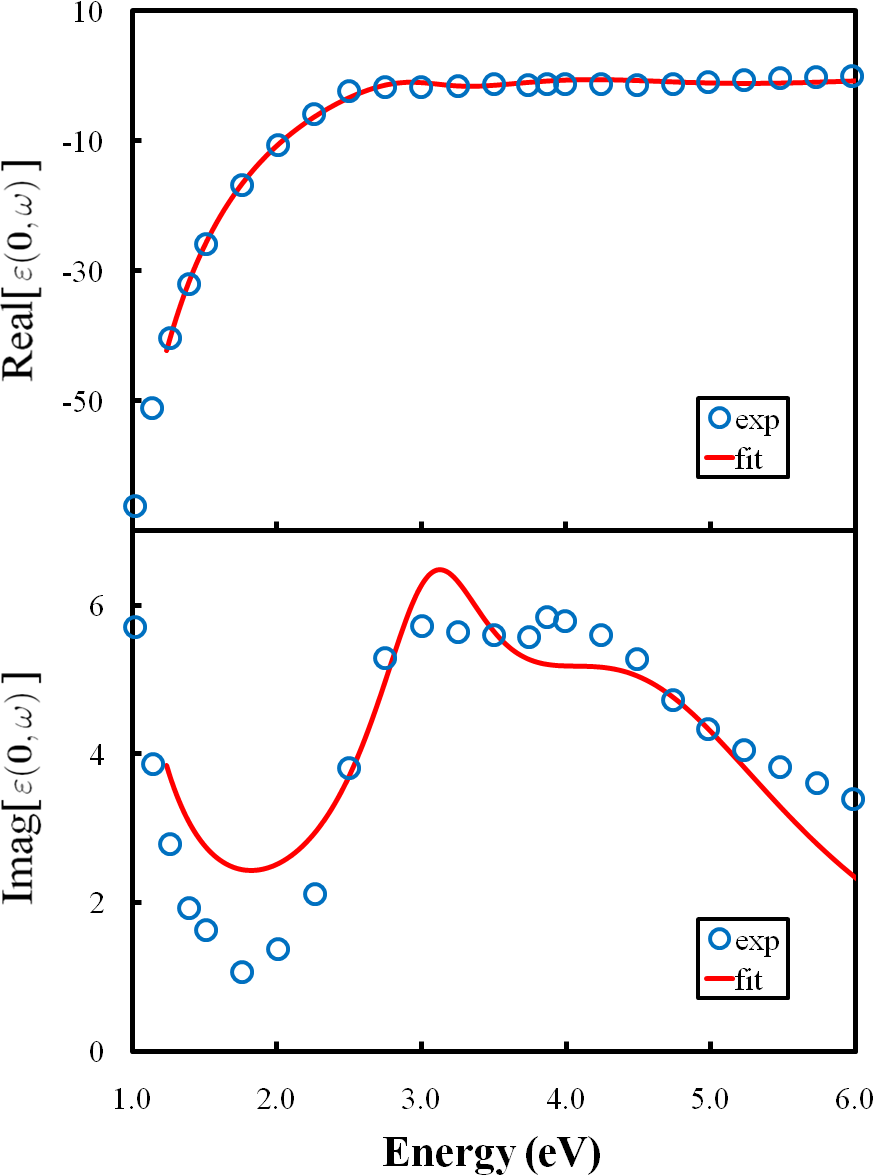}
  \caption{(color online) Fitted dielectric data for bulk Au (solid red lines), compared to that experimentally determined (open blue circles) \cite{OptConst_NobleMetals_JC}.}
  \label{fig:permittivity_Au_loc}
\end{figure}

A plot of calculated dielectric values against those experimentally determined is shown in Fig.\ \ref{fig:permittivity_Au_loc}. The fit is reasonably good given the simple form of Eq.\ (\ref{eq:electric_permittivity}). For example, features of the two interband transitions are captured near 3.15 and 4.30 eV, which is evident in Imag[$\varepsilon(\textbf{0}, \omega)$]. Note that $\omega_{\text{L}1}$ and $\omega_{\text{L}2}$ are also close to these values, as expected based on the discussion in Subsection \ref{sec:method:formulation}. Although, the fit is not as good as could be achieved with a more flexible function, such as an unrestricted fit. However, the present fitting scheme leads to parameters that are more physically realistic, and this is essential given that we are going to use these local ($\kvec = \textbf{0}$) parameters in the nonlocal ($\kvec \neq \textbf{0}$) expression. One consequence of this fit, for example, is that the minimum value of Imag[$\varepsilon(\textbf{0}, \omega)$] near 1.85 eV is not as small as the experimental one, which will end up giving broader plasmon near this energy resonances than expected. Fortunately, such differences will not play a significant role in the results that we present.

\begin{figure}
  \includegraphics[scale=0.6,bb=0 0 510 699]{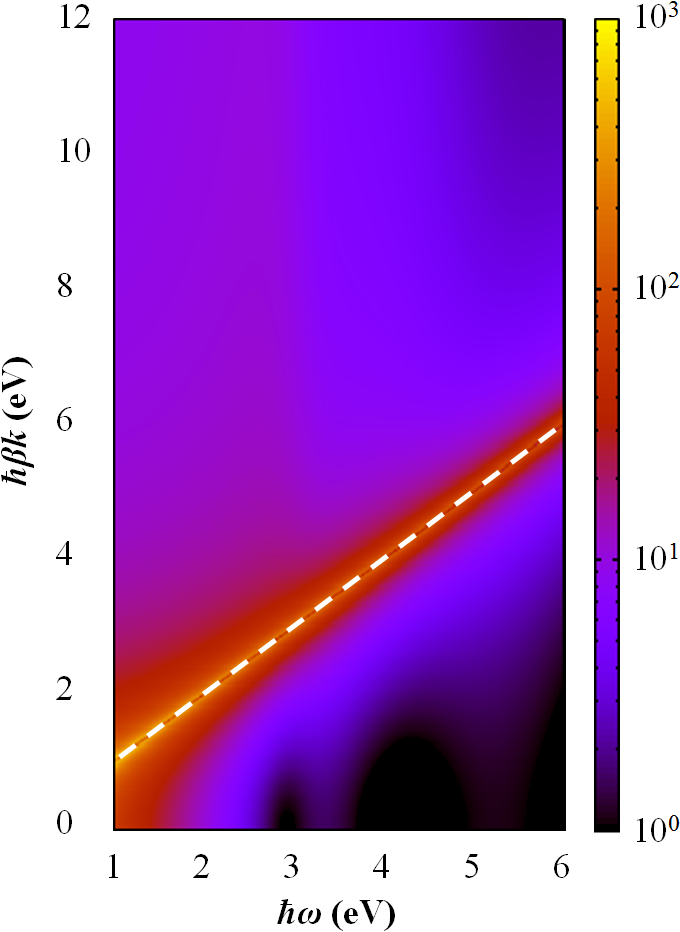}
  \caption{(color online) $|$Real$[\varepsilon(\kvec, \omega)]|$ of Au as a function of both $k$ and $\omega$. Below $\beta k \approx \omega$, $\varepsilon(\kvec, \omega) < 0$; and above, $\varepsilon(\kvec, \omega) > 0$. Note that no specific value is attached to $\beta^2$. The condition $\beta k = \omega$ is shown using a dashed white line.}
  \label{fig:permittivity_Au_NL}
\end{figure}

It is interesting to look at the evaluation of Eq.\ (\ref{eq:electric_permittivity}) as a function of both $k$ and $\omega$; Fig.\ \ref{fig:permittivity_Au_NL}.
Note that a slice through $k = 0$ gives the local dielectric data in Fig.\ \ref{fig:permittivity_Au_loc}.
When $\beta k \ll \omega$, $\varepsilon(\kvec, \omega)$ is relatively constant for a given $\omega$ -- i.e., it remains close to the local value. However, as $\beta k$ approaches $\omega$ from below, $\varepsilon(\kvec, \omega)$ quickly becomes very negative and changes sign rapidly as it passes through $\beta k \approx \omega$, after which $\varepsilon(\kvec, \omega)$ is thus no longer plasmonic. Absorption of light by a system is related to the value of $\varepsilon(\kvec, \omega)$ and the structure under consideration. For example, for a small spherical particle in air, the maximum absorption occurs when Real$[\varepsilon(\kvec, \omega)] = -2$ \cite{Light-Scattering_BH}. Figure \ref{fig:permittivity_Au_NL} therefore indicates that in addition to the local absorption, additional (anomalous) absorption will occur when $\beta k \approx \omega$ \{when the rapid variation in Real$[\varepsilon(\kvec, \omega)]$ occurs\}.

Nonlocal effects are very prominent for small structures \cite{McMahon_NLDiel}, as we will demonstrate below. In such systems, it is necessary to consider the reduced mean free path of the \textit{sp}-band electrons due to electron--interface scattering. As was briefly discussed in Section \ref{sec:intro}, this can be taken into account by using a modified collision frequency in Eq.\ (\ref{eq:hydrodynamic-Drude}) \cite{Schatz_size-dep-damping}: $\gamma' = \gamma + A v_\text{F} / L_\text{eff}$, where the effective mean free electron path is $L_\text{eff} = 4 V / S$ in 3D and ${\pi} S / P$ in 2D, where \textit{V} is the volume of the structure with surface area $S$ with perimeter $P$, 
and $A$ can be considered the proportion of electron--interface collisions that are totally inelastic. Such scattering can also be considered a nonlocal effect \cite{A-param_Apell_PRL_1983, Abajo_2008}. In a formal sense, $A$ is related to the translational invariance at the surface, the full description of which depends on the geometry and morphology of the structure, its local dielectric environment \cite{A-param_Pinchuk_2004}, and the dielectric function of the material, which is ultimately nonlocal in character. Although, the general magnitude of $A$ can be arrived at in the local limit, and in a variety of ways \cite{Opt-Prop_Clusters_Vollmer}. Because of these complex details, correctly choosing the value of $A$ can be challenging, and large values can have a significant effect \cite{McMahon_NL_nanoshells}. Therefore, for simplicity, and consistency with our previous Letter \cite{McMahon_NLDiel}, we take $A = 0.1$ for the calculations herein.

\section{Computational Considerations}
\label{sec:comp_consid}

Computational domains were discretized using a Yee spatial lattice \cite{Yee_FDTD}, as outlined in the Appendix. The edges of the domains were truncated using convolutional perfectly matched layers \cite{CPML_Gedney}. For the calculations in Section \ref{sec:rslts:subsec:verification}, variable grid spacings were used for this discretization, as will be discussed, as well as the low-frequency 2D value of $\beta^2$ (for consistency with our previous Letter \cite{McMahon_NLDiel}). For the calculations on metal films in Subsection \ref{sec:rslts:apps:films}, grid spacings of 0.1 nm were used for the 2-nm film, and 0.2 nm for the others, as well as the high-frequency 2D value of $\beta^2$. For the nanoparticles in Subsection \ref{Section:Rslts_Nanoparticles}, grid spacings of 0.2 nm were used for the 4 and 7-nm nanoparticles, and 0.5 nm for the 15-nm one. For all of the nanowires in Subsection \ref{sec:rslts_nanowires}, grid spacings of 0.25 nm were used. In both of these latter cases, the high-frequency values of $\beta^2$ were used.

Optical responses were determined by calculating extinction cross sections \cite{Light-Scattering_BH} (the amount of power absorbed or scattered relative to the incident light), by integrating the normal component of the Poynting vector around surfaces enclosing the particles \cite{Gray_optoelectronics_2003} \footnote{We have determined that an overall increase in the cross section can occur by performing the integration too close to the structure. In our previous Letter \cite{McMahon_NLDiel}, this resulted in the cross sections for the 2-nm cylindrical nanowire and 5-nm triangular one to approach 0.1 at low energies, rather than 0.0.}. In order to obtain accurate Fourier-transformed fields necessary for such calculations, as well as field intensity profiles, incident Gaussian damped sinusoidal pulses with frequency content over the range of interest (1 -- 6 eV) were introduced into the computational domains using the total-field--scattered-field technique \cite{TFSF_Taflove}. Furthermore, all simulations were carried out to at least 100 fs.

Before leaving this subsection, we mention that instabilities have been encountered in some 3D calculations. For example, simulations of 1.0-nm Au nanoparticles become unstable when grid spacings of 0.05 nm are used. However, in 2D, such instabilities do not seem to exist. We are currently investigating this issue.

%
%
\section{Numerical Verification}
\label{sec:rslts:subsec:verification}

One way to determine the accuracy of the method presented in Subsection \ref{sec:method:formulation} and the Appendix is to compare computed results to obtainable analytical ones. Such comparisons are possible for metal films \cite{Fuchs_metal-films_1969}, cylindrical wires \cite{nl_cylinder_analytic_Ruppin_1989}, and spherical particles \cite{nl_diel-sphere_analytic_Ruppin_1981}. In this section, such a comparison is made for a $r = 2$ nm radius cylindrical nanowire, an example that we considered in our previous Letter \cite{McMahon_NLDiel}.

\begin{figure}
  \includegraphics[scale = 0.35, bb=0 0 934 645]{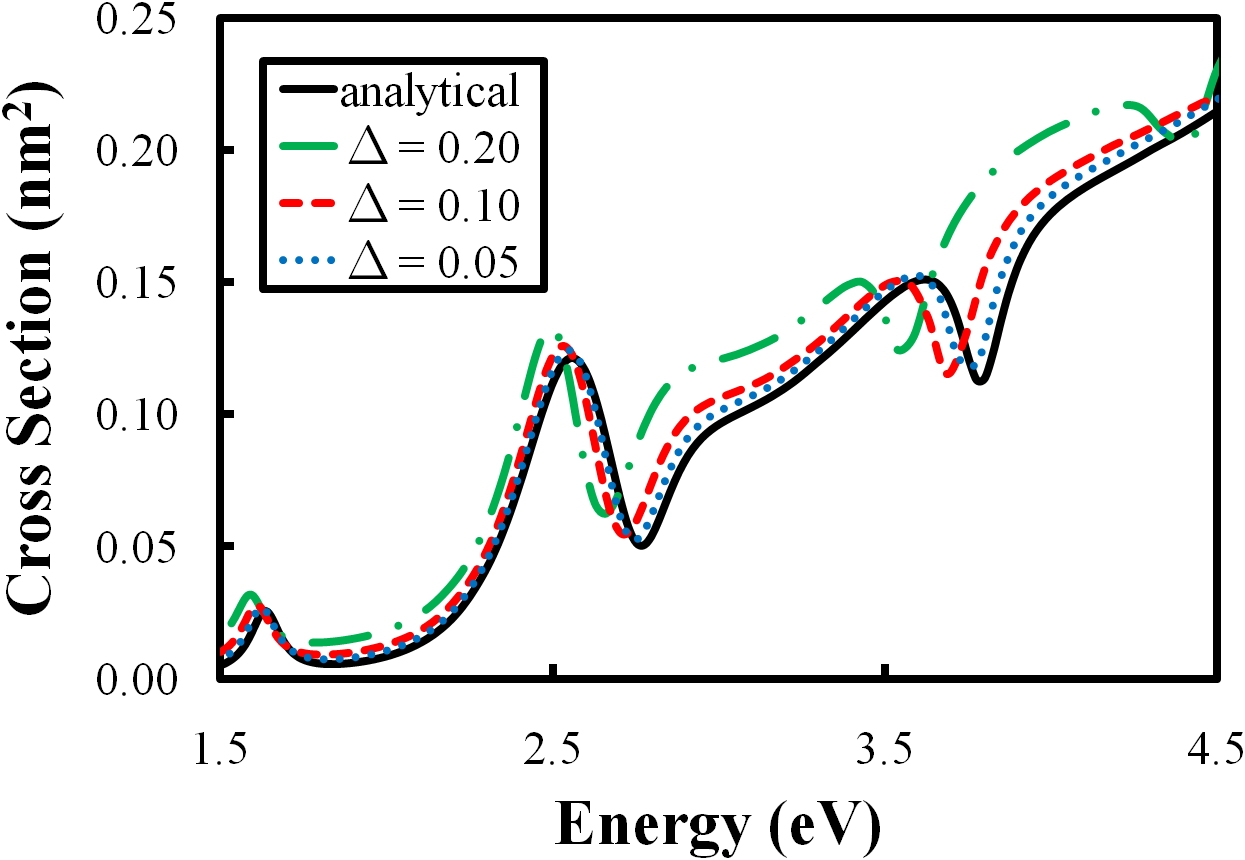}
  \caption{(color online) Convergence of calculated extinction cross sections to the analytical result \cite{nl_cylinder_analytic_Ruppin_1989} for a 2 nm radius cylindrical Au nanowire, with respect to the grid spacing $\Delta$.}
  \label{fig:r2nm-cyl_analyt_compare_diff-dx}
\end{figure}

The optical responses calculated using uniform grid spacings $\Delta$ in both $x$ and $y$ of $0.2$, $0.1$, or $0.05$ nm are compared to the analytical result \cite{nl_cylinder_analytic_Ruppin_1989} in Fig.\ \ref{fig:r2nm-cyl_analyt_compare_diff-dx}.
A number of peaks and valleys are seen in all results, which correspond to the dipolar LSPR near 2.55 eV and anomalous absorption near 1.61, 2.75, and 3.78 eV. A discussion of these effects was given in Ref.\ \onlinecite{McMahon_NLDiel}, and will be further elaborated on below. The calculated and analytical results are found to agree remarkably well, both qualitatively and quantitatively, providing numerical verification of our method. Additionally, decreasing the grid spacing leads to significantly better results, especially for the higher energy peaks. For example, the additional peak near 3.78 eV converges from 3.53 to 3.68 to 3.72 eV as $\Delta$ is reduced from $0.2$ to $0.1$ to $0.05$ nm, respectively. Such convergence is understandable, because for a given grid spacing, there is an uncertainty in $r$ of $\pm \Delta$. Since the results appear to redshift with increasing $\Delta$, we can infer that the nanowire radius is approximately $r + \Delta$. The convergence of these effects is much tougher than those in local electrodynamics, where even $\Delta = 0.2$ nm is sufficient (not shown). These results demonstrate the exquisite sensitivity of nonlocal effects to even minor geometric features.

\begin{figure}
  \includegraphics[scale = 0.27, bb=0 0 934 1636]{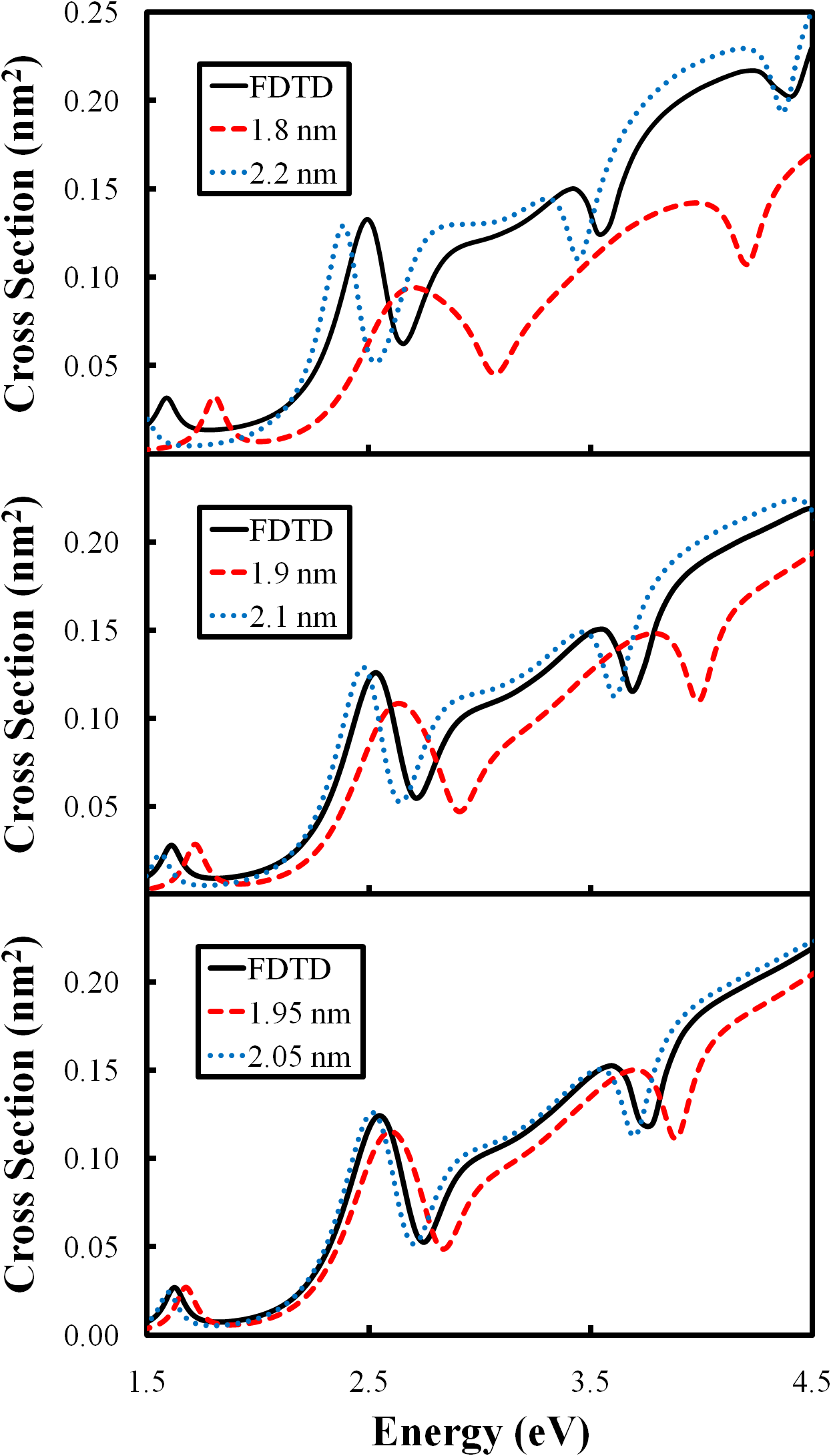}
  \caption{(color online) Extinction cross sections of 2 nm radius cylindrical Au nanowires, calculated using grid spacings of $\Delta = $ (top) $0.2$, (middle) $0.1$, and (bottom) $0.05$ nm. The results show that the calculations (solid black lines), denoted as FDTD, are always constrained by the analytical results (broken lines) when tolerances for the grid spacings are considered.}
  \label{fig:cyl_analyt_dx-tol}
\end{figure}

Grid spacings of $0.05$ or $0.1$ nm are impractical for most calculations, because of the resulting computational effort required. At first this appears troublesome, given that the nonlocal effects are so sensitive to this parameter. However, the only real downside is that an uncertainty in the geometry of $\pm \Delta$ must be accepted (which is also the case in local electrodynamics, but is less important). This is because it is found that for a given grid spacing, the calculated results always fall between the analytical ones that incorporate the $\pm \Delta$ tolerance; see Fig.\ \ref{fig:cyl_analyt_dx-tol}.
For example, in the case of a $r = 2$ nm cylindrical nanowire, the calculated results with $\Delta = 0.2$ are constrained by the analytical ones with $r = 1.8$ and $2.2$ nm. Results for $\Delta = 0.1$ and $0.05$ nm are shown in Fig.\ \ref{fig:cyl_analyt_dx-tol} as well, and are also consistent with this analysis.

\section{Applications }
\label{sec:rslts:apps}

%
%
\subsection{Metal Films (1D Systems)}
\label{sec:rslts:apps:films}

\begin{figure}
  \includegraphics[scale = 0.35, bb=0 0 1239 440]{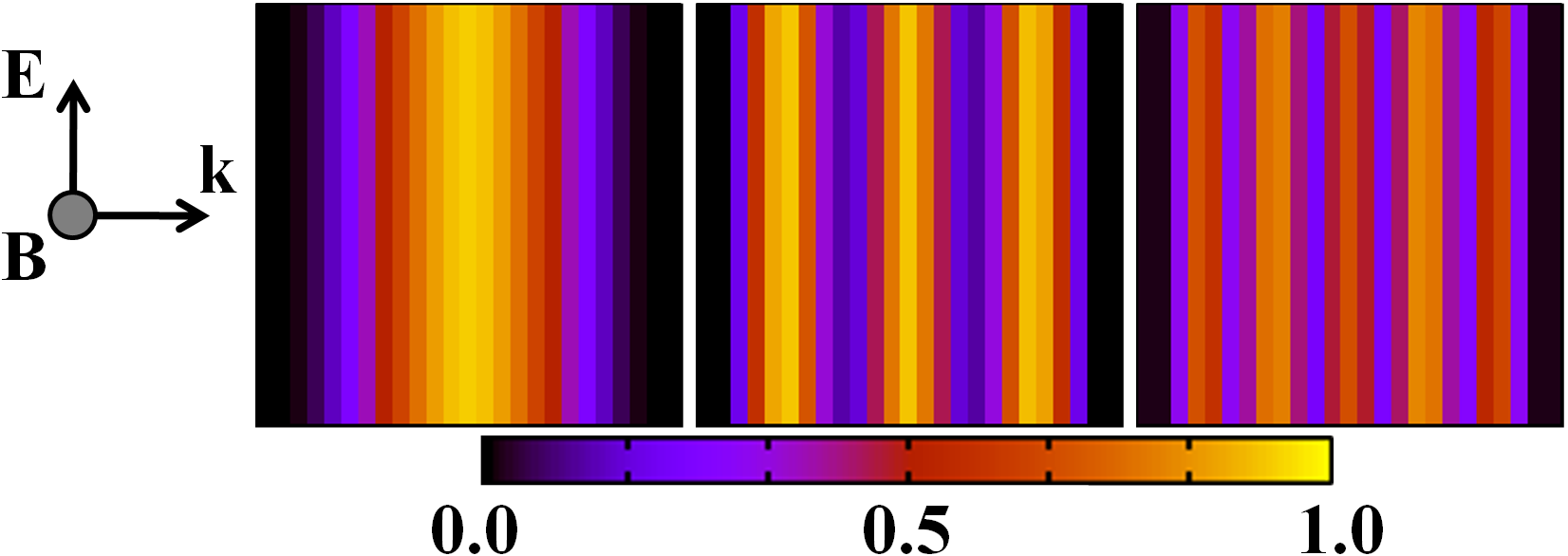}
  \caption{(color online) Normalized $|\textbf{D}|^{2}$ intensity profiles inside a 2 nm thick Au film at energies of (left) 1.14, (middle) 3.36, and (right) 5.54 eV. The polarization and direction of incident light is indicated; in each image, the sides of the metal film are padded on the left and right using solid black lines.}
  \label{fig:D2_film}
\end{figure}

In this subsection, we study the transmission, reflection, and absorption of thin Au films illuminated at normal incidence (see Fig.\ \ref{fig:D2_film} for a schematic diagram of the incident polarization), which have an effective dimension of one. For simplicity, we take the surrounding medium to be air. Although, it would be straightforward to introduce other dielectric layers into the calculations. Because wavevector components only exist for the direction normal to the surface, these systems are ideal for studying and qualitatively highlighting nonlocal effects. Furthermore, they allow us to draw some connections with related experimental results \cite{NL-exp_Fitton_1971, NL-exp_Lindau_1970}.

\begin{figure}
  \includegraphics[scale=0.4, bb=0 0 622 1086]{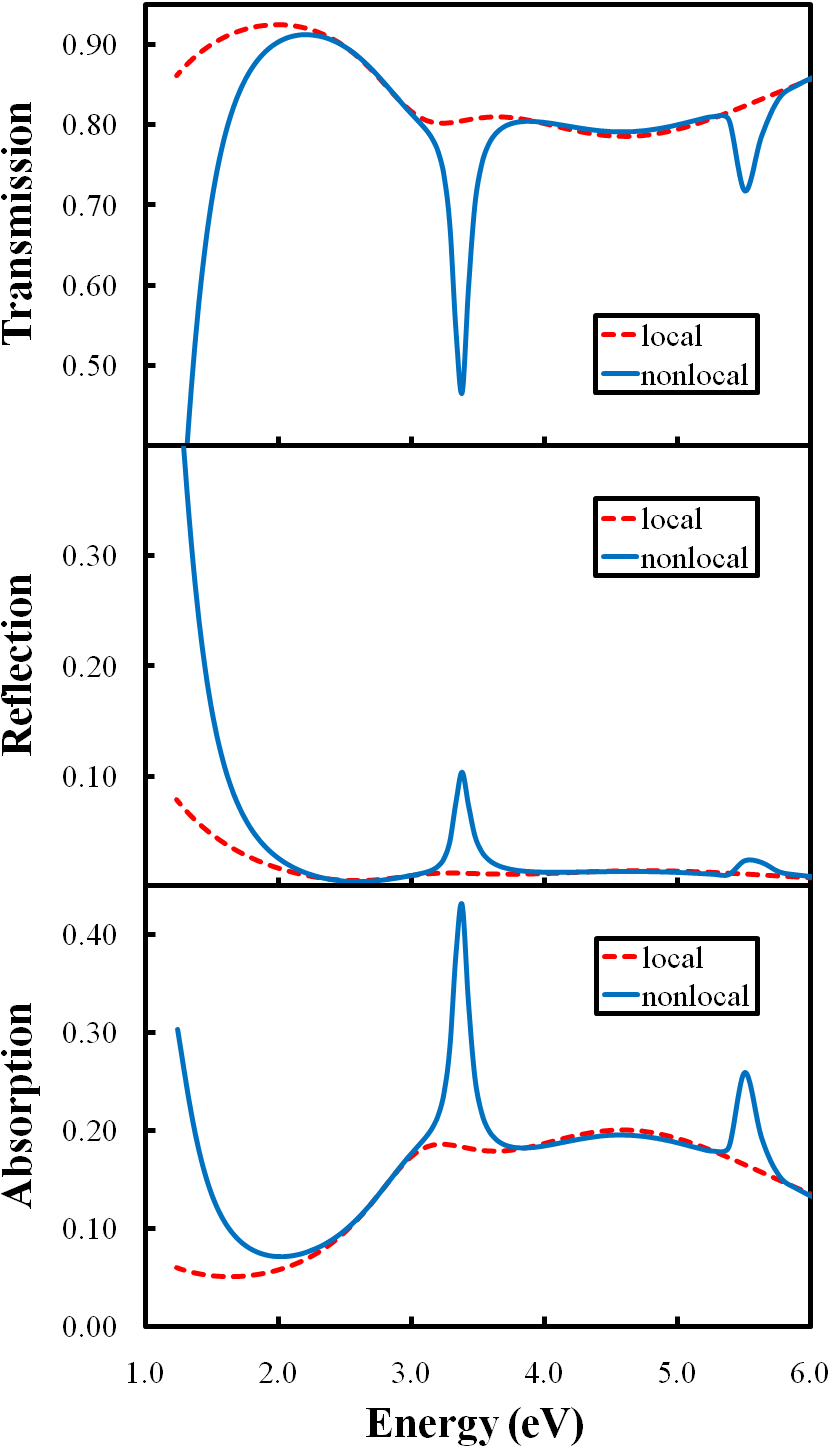}
  \caption{(color online) (top) Transmission, (middle) reflection, and (bottom) absorption spectra for a 2 nm thick Au film illuminated at normal incidence. Both local (broken red lines) and nonlocal (solid blue lines) calculations are shown.}
  \label{fig:film_2nm}
\end{figure}

\begin{figure}
  \includegraphics[scale=0.3, bb=0 0 836 1460]{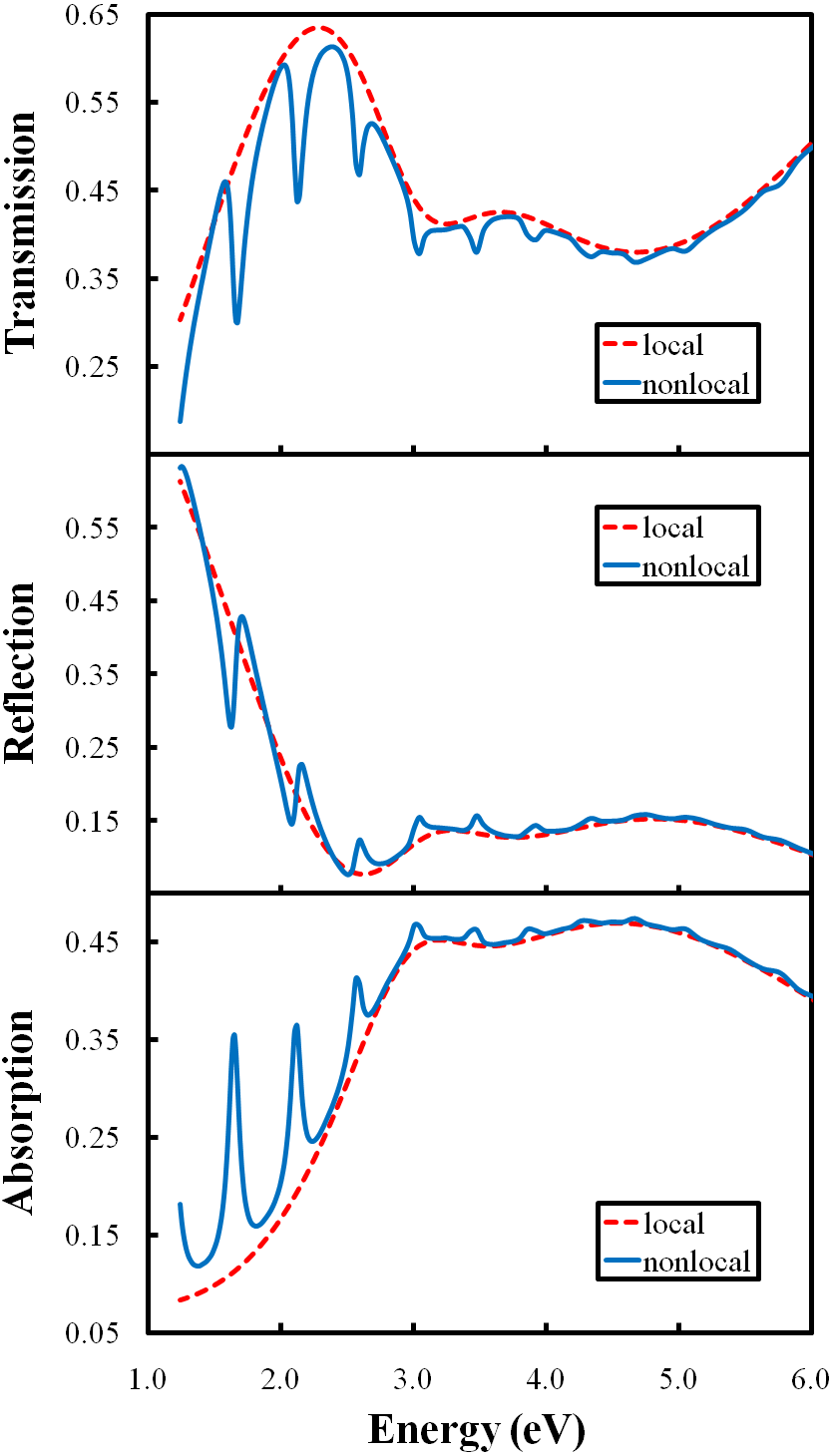}
  \caption{(color online) (top) Transmission, (middle) reflection, and (bottom) absorption spectra for a 10 nm thick Au film illuminated at normal incidence. Both local (broken red lines) and nonlocal (solid blue lines) calculations are shown.}
  \label{fig:film_10nm}
\end{figure}

\begin{figure}
  \includegraphics[scale=0.3, bb=0 0 834 1452]{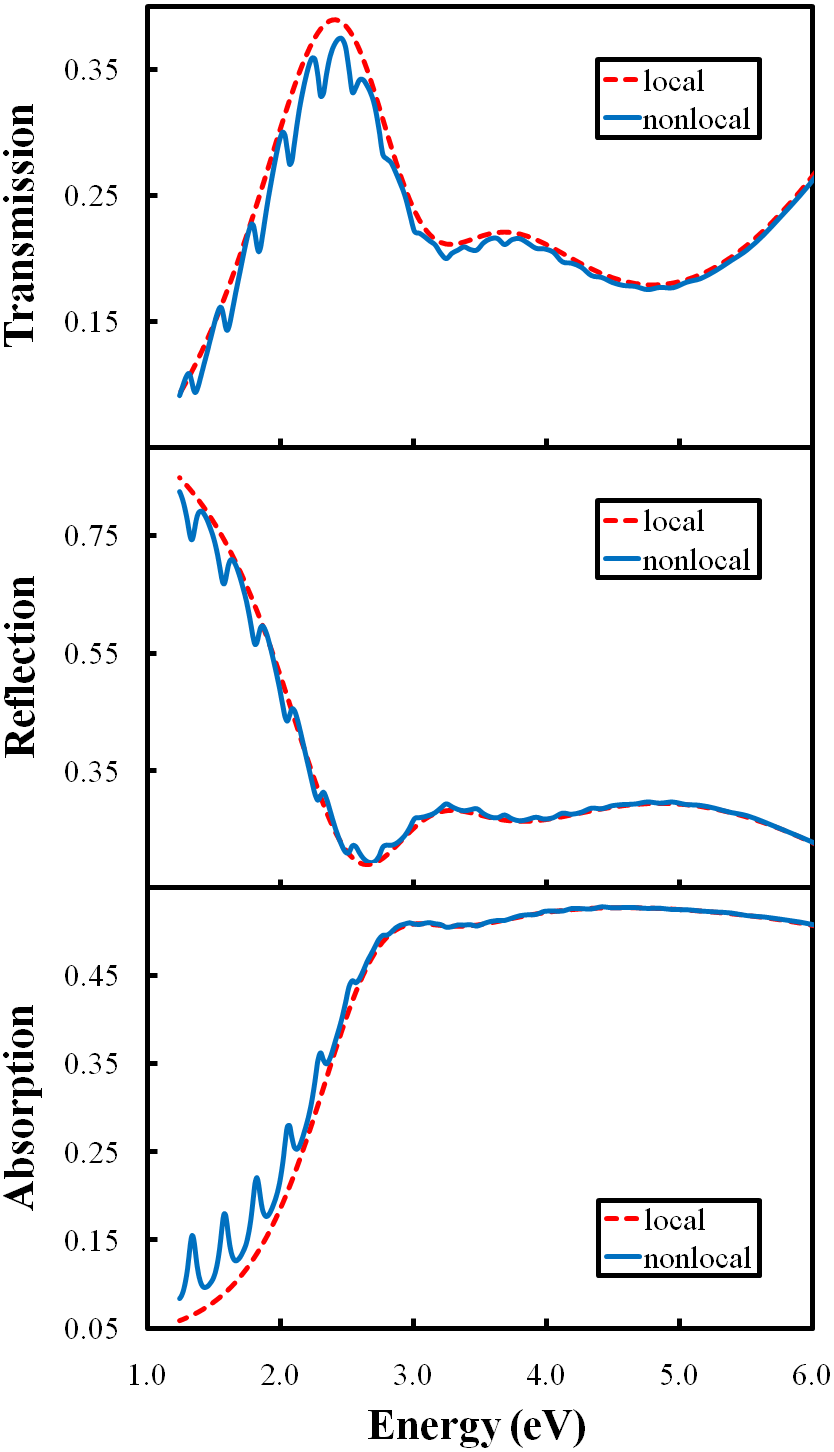}
  \caption{(color online) (top) Transmission, (middle) reflection, and (bottom) absorption spectra for a 20 nm thick Au film illuminated at normal incidence. Both local (broken red lines) and nonlocal (solid blue lines) calculations are shown.}
  \label{fig:film_20nm}
\end{figure}

The transmission, reflection, and absorption spectra for 2, 10, and 20 nm thick films are shown in Figs.\ \ref{fig:film_2nm} -- \ref{fig:film_20nm}, respectively.
In the absorption spectra, narrow additional (anomalous) absorption peaks occur in the nonlocal results, relative to the local ones. The appearance of these peaks is identical to theoretical predictions \cite{Fuchs_metal-films_1969} and experimental observations \cite{NL-exp_Fitton_1971, NL-exp_Lindau_1970} on other thin metal films, where they are the result of optically excited longitudinal (or volume) plasmons. (The effects are called such because they are longitudinal to $\kvec$ and are contained within the volume of the structure, unlike surface plasmons, which propagate along a metal--dielectric interface.) Not surprisingly, at the anomalous absorption energies, there is corresponding decrease in the transmission. However, contrary to the initial expectation of a decrease in reflection, we find that there can be either an increase or a decrease, depending on if the corresponding absorption occurs well above (giving an increase) or below (giving a decrease) the surface plasmon energy, which is around 2.65 eV for the 10-nm film, for example.

Although a little hard to discern from Figs.\ \ref{fig:film_2nm} -- \ref{fig:film_20nm}, but can be inferred from previous results \cite{McMahon_NLDiel}, the anomalous absorption peaks redshift as the film thickness is increased. This causes many more such peaks that were at higher energies to appear in the optical range. For example, there are three peaks for the 2-nm film (Fig.\ \ref{fig:film_2nm}), but twelve for the 10-nm one (Fig.\ \ref{fig:film_10nm}). In addition, their intensities drastically decrease, where by 20-nm, the nonlocal results are almost converged to the local ones; Fig.\ \ref{fig:film_20nm}. We will revisit these points below.

In order to determine whether the anomalous absorption in these results is actually from the excitation of longitudinal plasmons, intensity profiles of $|\textbf{D}|^{2}$ at the anomalous absorption energies for the 2-nm film [1.14 (not shown in Fig.\ \ref{fig:film_2nm}), 3.36, and 5.54 eV] can be examined; Fig.\ \ref{fig:D2_film}. 
Well-defined standing-wave patterns of $|\textbf{D}|^{2}$ are seen inside the films longitudinal to $\kvec$, confirming the assumption of their nature. The wavelengths of these standing waves is found to satisfy the condition
\begin{equation}
  \label{eq:lambda_L}
  \lambda_\text{L} = 2 d / m ~~ ,
\end{equation}
where $d$ is the film thickness and $m = 1,3,5,...$, which means that odd numbers of half-wavelengths fit longitudinally into the structures. The wavelengths defined by Eq.\ (\ref{eq:lambda_L}) will hereon be referred to as ``modes'', characterized by $m$. In Fig.\ \ref{fig:D2_film}, it is the $m = 1,3,$ and $5$ modes that are explicitly shown. These results are significantly different from the local ones, where relatively uniform $|\textbf{D}|^{2}$ patterns are found at all energies (not shown). Nonetheless, as mentioned above, this analysis agrees with previous results on analogous systems \cite{Fuchs_metal-films_1969, NL-exp_Fitton_1971, NL-exp_Lindau_1970}, providing further support for the validity of our method. Based on the observations in Fig.\ \ref{fig:D2_film} and the analysis above, it makes sense that the anomalous absorption features should redshift with increasing thickness and that their intensity should decrease with increasing $m$.

From the discussion above, and that in Subsection \ref{Section:Diel_Model}, the approximate anomalous absorption energies can be predicted. From Eq.\ (\ref{eq:hydrodynamic-Drude}), it is seen that rapid variations in $\varepsilon(\kvec, \omega)$ will occur when $\omega \approx \beta k$, which will likely lead to an absorption condition; and from Eq.\ (\ref{eq:lambda_L}), it is seen that longitudinal plasmons with wavelength $\lambda_\text{L}$ are excited inside a structure of thickness $d$, which will result in momentum states with magnitude $k = 2 \pi / \lambda_\text{L}$. Thus, everything needed to predict the approximate anomalous absorption energies is known: $\hbar \omega$ = $m \beta \pi / d$. Using the 2-nm film as an example, this analysis predicts anomalous absorption at energies of $\hbar \omega = m \cdot 1.44$ eV. For the first three $m$ modes, these are 1.44, 4.31, and 7.19 eV, while those rigorously calculated are 1.14, 3.36, and 5.54 eV, respectively; Fig.\ \ref{fig:film_2nm}. While not exact, the predictions are reasonably close. Part of these differences can be attributed to the grid spacing error, as outlined in Section \ref{sec:rslts:subsec:verification}, which in this case leads to an uncertainty in $d$ of $\pm 0.2$ nm. This analysis could also be applied to related experimental results \cite{NL-exp_Fitton_1971}. However, it is important to keep in mind that this is a simple approximation, and if appropriate, more accurate values should be obtained from full calculations or, if possible, by using rigorous theory \cite{Fuchs_metal-films_1969}.

\subsection{Spherical Nanoparticles (3D Systems)}
\label{Section:Rslts_Nanoparticles}

\begin{figure}
  \includegraphics[scale=0.27, bb=0 0 910 1637]{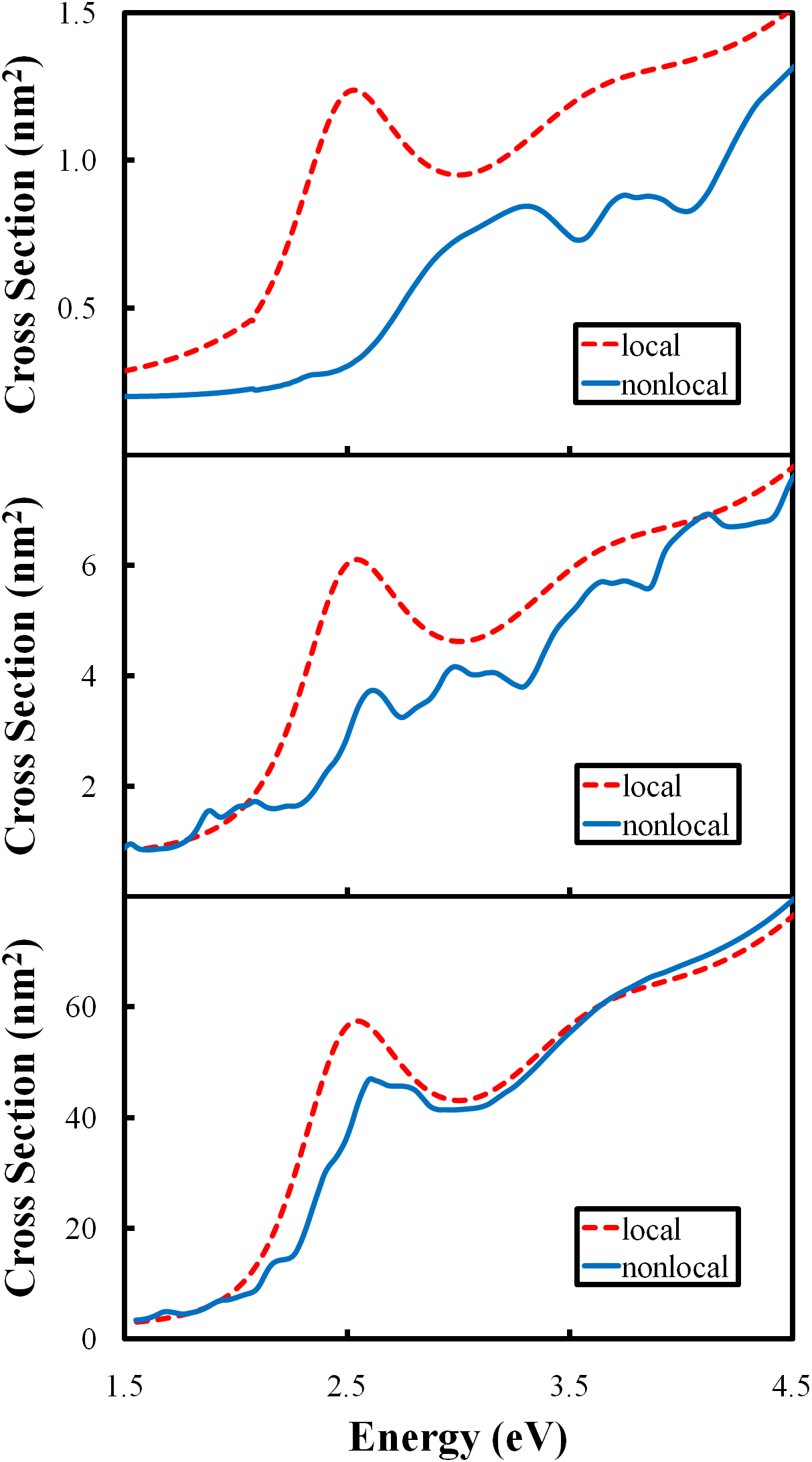}
  \caption{(color online) Extinction cross sections of spherical Au nanoparticles with diameters of (top) 4, (middle) 7, and (bottom) 15 nm. Both local (broken red lines) and nonlocal (solid blue lines) calculations are shown.}
  \label{fig:Au_nanoparticles}
\end{figure}

In this subsection, we study spherical nanoparticles, utilizing the full 3D nonlocal electrodynamics method outlined in Subsection \ref{sec:method:formulation} and the Appendix. The optical responses of nanoparticles with diameters of 4, 7, and 15 nm are shown in Fig.\ \ref{fig:Au_nanoparticles}. (The polarization and direction of incident light is irrelevant.) Note that for small nanoparticles, such as these, the optical responses are predominately absorption, as scattering does not play a significant role for sizes less than approximately 20 nm.
Figure \ref{fig:Au_nanoparticles} shows that inclusion of nonlocal effects results in significant anomalous absorption and LSPR blueshifting for both the 4 and 7-nm nanoparticles. In fact, these effects are so large that the main LSPRs are hardly even distinguishable. 

As discussed in Subsection \ref{sec:rslts:apps:films}, the anomalous absorption peaks arise from the excitation of longitudinal plasmons. However, unlike the systems discussed in that subsection, this effect diminishes much faster as the nanoparticle size is increased, such that at 15-nm, the anomalous peaks show up only as slight indents on the main LSPR. These differences can be attributed to two effects. First of all, in a spherical nanoparticle, scattering of the incident light off of the exterior surface generates many $\kvec$-components that can interact and dephase one another, especially for the high-order $m$ modes with multiple nodes. Secondly, scattering of the conduction electrons that compose the longitudinal plasmons off of the interior surface can also lead to dephasing. Each of these processes causes nonlocal effects to diminish at much smaller distances in spherical nanoparticles, relative to metal films.

LSPR blueshifting is most apparent for the 15-nm nanoparticle. This is simply because the anomalous absorption is low, allowing this peak to be clearly identified. The local LSPR is seen at 2.57 eV, while the nonlocal one is seen at 2.71 eV. This effect can be understood by looking at the form of Eq.\ (\ref{eq:hydrodynamic-Drude}). When nonlocal effects are included, the interplay between $\omega$ and $\kvec$ causes all effects \{e.g., the absorption condition of Real$[\varepsilon(\kvec, \omega)] = -2$, in this case\} to appear at higher energies compared to $\kvec = \textbf{0}$.

Based on the results in Fig.\ \ref{fig:Au_nanoparticles}, one might wonder why such strong nonlocal effects have not been experimentally observed in such systems. Obviously these effects are important, and have been observed in other cases \cite{NL-exp_Lindau_1970, NL-exp_Fitton_1971}. There are many possible reasons for this. The most probable one is that experimental measurements are often made on heterogeneous collections of nanoparticles. Given that nonlocal effects are very sensitive to nanoparticle dimensions (see Section \ref{sec:rslts:subsec:verification}, for example), slight heterogeneity could effectively average them away. Support for this claim comes from an experimental study of isolated Au nanoparticles \cite{Novotny_2006}, which clearly demonstrated the LSPR blueshift, and possibly anomalous absorption features \cite{McMahon_NLDiel}. Another possible explanation is that our choice of $\beta^2$ is not optimal (which is directly related to the strength of nonlocal effects), as we have recently argued for metallic nanoshells \cite{McMahon_NL_nanoshells}. The hydrodynamic Drude model neglects quantum mechanical exchange and correlation effects, which in a local density approximation would decrease $\beta^2$. A third possible explanation is that our choice of damping parameter $A$ is too low. Increasing this would smooth all spectral features (i.e., the anomalous absorption would not appear as strong). Support for this comes from a combined theoretical (local electrodynamics) and experimental study of metallic nanoshells, where values of greater than 1.0 are needed to describe the results (this corresponds to $L_\text{eff}$ reduced below that based on geometric considerations alone) \cite{Halas_nanoshells_lineshape}.

\subsection{Nanowires (2D Systems)}
\label{sec:rslts_nanowires}

In our previous Letter \cite{McMahon_NLDiel}, we demonstrated that nonlocal effects are particularly important in structures with apex features, such as triangular nanowires. In such structures, optical responses can be affected by nonlocal effects for much larger sizes than for those with smooth geometries, such as cylindrical nanowires. Additionally, even though far-field optical properties seem to converge to local ones at large sizes with regard to anomalous absorption and LSPR blueshifting (to a large extent), the near-field properties, such as electric field enhancements, hereon referred to as $| \textbf{E} |^2$ enhancements, do not.

As a final application of the method presented in Subsection \ref{sec:method:formulation} and the Appendix, we study the optical responses and $| \textbf{E} |^2$ enhancements around isolated cylindrical, square, and triangular nanowires with 50 nm diameters or side-lengths, common sizes used in experimental and theoretical studies. $| \textbf{E} |^2$ enhancements from such structures have been studied in the past \cite{Martin_nanowires_2000, OJFMartin_Ag_nanowires_PRB-2001}. However, to the best of our knowledge, all previous studies have been carried out using local electrodynamics (at least for non-cylindrical structures), except for the aforementioned discussion in our previous Letter \cite{McMahon_NLDiel}.

\begin{figure}
  \includegraphics[scale=0.3, bb=0 0 810 1452]{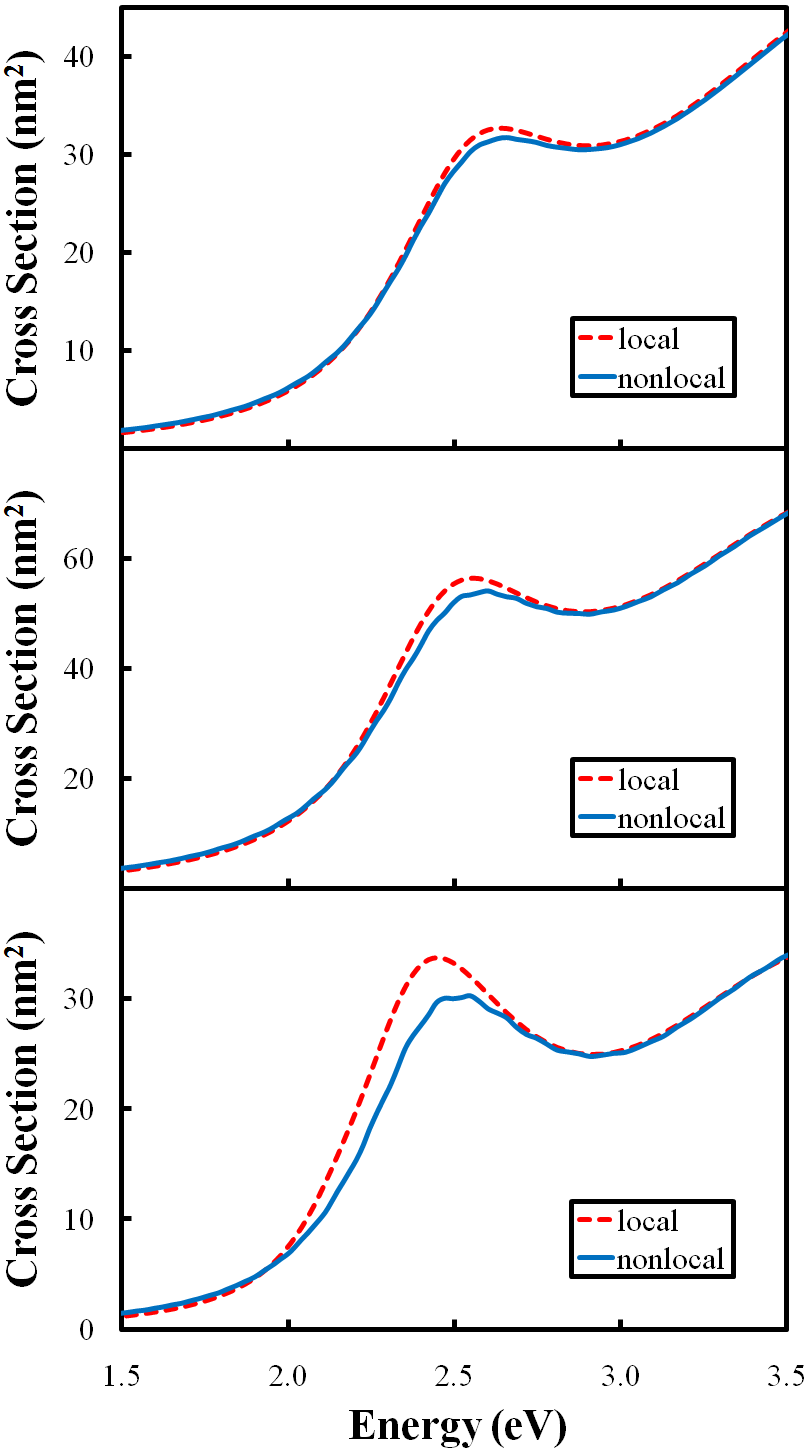}
  \caption{(color online) Extinction cross sections of (top) cylindrical, (middle) square, and (bottom) triangular Au nanowires with diameters or side-lengths of 50 nm. Each was illuminated with the $\textbf{E}$-field polarized along the longest axis of the structure. Both local (broken red lines) and nonlocal (solid blue lines) calculations are shown.}
  \label{fig:nanowires_opt-resp}
\end{figure}

\begin{figure}
  \includegraphics[scale=0.3, bb=0 0 1368 725]{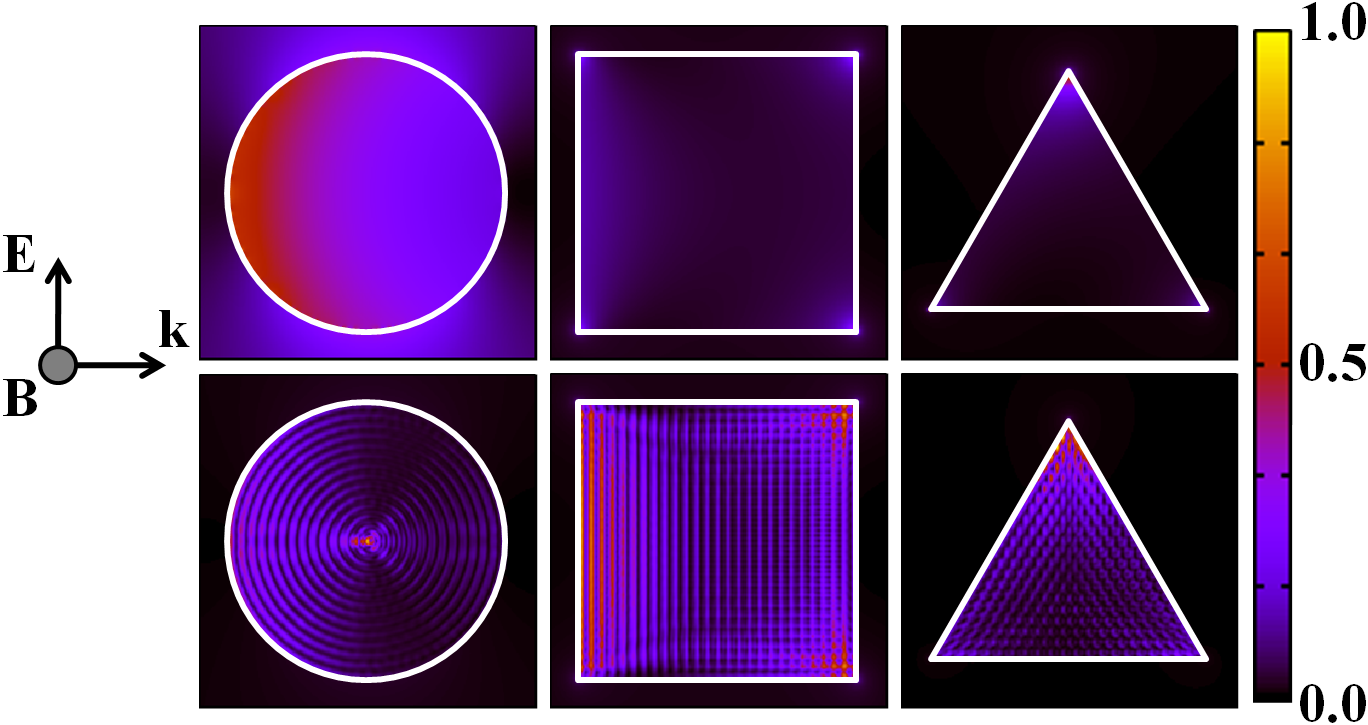}
  \caption{(color online) Normalized $|\textbf{D}|^{2}$ intensity profiles at the LSPR energies in and around (left) cylindrical, (middle) square, and (right) triangular Au nanowires with diameters or side-lengths of 50 nm. Both (top) local and (bottom) nonlocal calculations are shown. The polarization and direction of incident light is indicated; the nanowires are outlined in white.}
  \label{fig:nanowires_D2}
\end{figure}

\begin{figure}
  \includegraphics[scale=0.3, bb=0 0 1368 725]{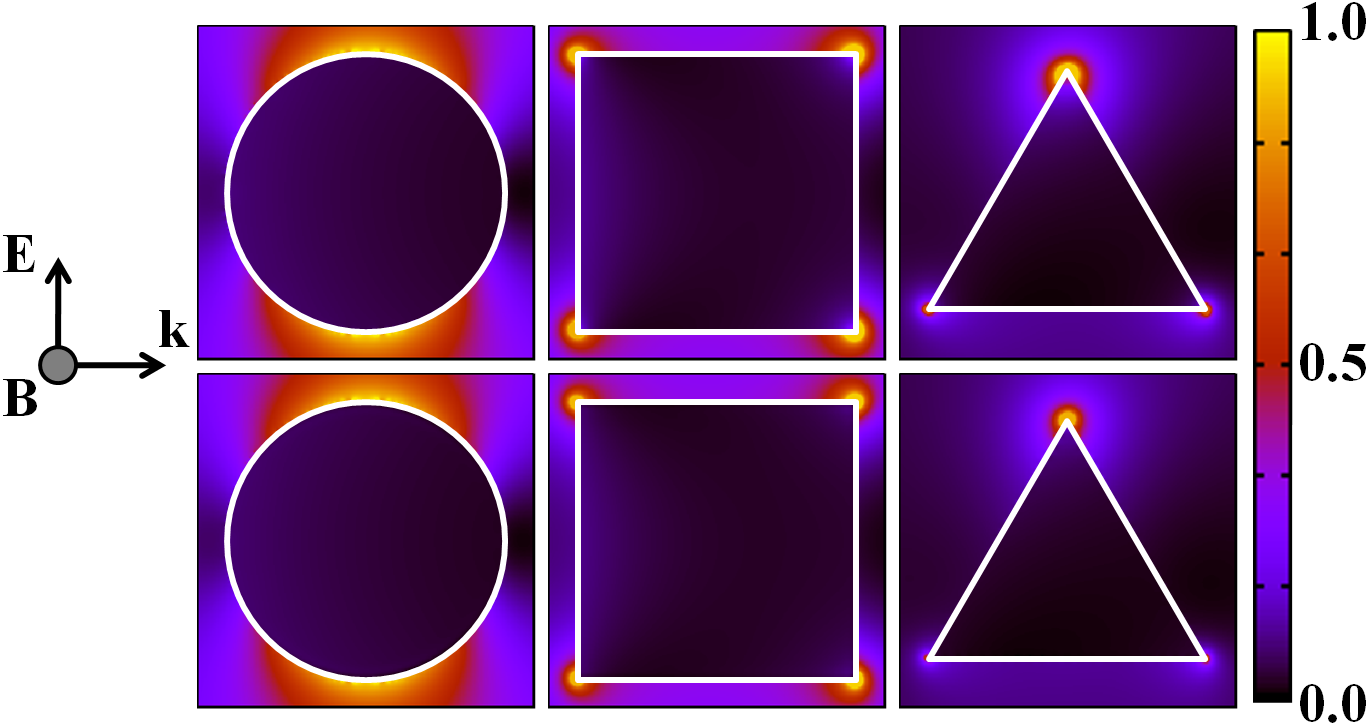}
  \caption{(color online) Normalized $|\textbf{E}|^{2}$ intensity profiles at the LSPR energies in and around (left) cylindrical, (middle) square, and (right) triangular Au nanowires with diameters or side-lengths of 50 nm. Both (top) local and (bottom) nonlocal calculations are shown. The polarization and direction of incident light is indicated; the nanowires are outlined in white.}
  \label{fig:nanowires_E2}
\end{figure}

Optical responses of nanowires illuminated with the $\textbf{E}$-field polarized along the long axis of each structure are shown in Fig.\ \ref{fig:nanowires_opt-resp}. (Schematic diagrams of the polarization are shown in Figs.\ \ref{fig:nanowires_D2} and \ref{fig:nanowires_E2}.) 
Because of their large sizes, these structures do not exhibit distinct anomalous absorption. Nonetheless, many closely spaced longitudinal plasmon modes do exist (\textit{vide infra}), which leads to very minor, closely spaced ``bumps'' in the nonlocal results, as well as LSPR blueshifting. These modes can again be confirmed by looking at intensity profiles of $| \textbf{D} |^2$; Fig.\ \ref{fig:nanowires_D2}.
Unlike the results in Fig.\ \ref{fig:D2_film}, the longitudinal plasmons in Fig.\ \ref{fig:nanowires_D2} form much more complex patterns. These can be attributed to two (related) effects. One is that the size of the structure along the longitudinal direction of the incident field is not the same at all positions. Therefore, for a given energy, modes of different order will be sustained at multiple positions along the structure, at each place where Eq.\ (\ref{eq:lambda_L}) is satisfied \cite{McMahon_NLDiel}. This is also one of the reasons why nonlocal effects are so strong in structures with apex features, and why they can remain important in them for arbitrarily large sizes. In other words, low-order longitudinal plasmon modes can always be sustained near the apex. And two, scattering of the incident field off of a curved nanowire surface generates many $\kvec$-components, which can excite longitudinal plasmons along directions other than that of the incident $\kvec$, creating an interference pattern. This effect also leads to the dephasing of longitudinal plasmons, as discussed in Subsection \ref{Section:Rslts_Nanoparticles}.

It is also interesting to look at intensity profiles of $| \textbf{E} |^2$ at the LSPR energies, near where this quantity is expected to be maximized \cite{SERS_T-rod_VanDuyne_2008}; Fig.\ \ref{fig:nanowires_E2}.
(Note that these energies are slightly different in the local and nonlocal results, due to LSPR blueshifting.) In Fig.\ \ref{fig:nanowires_E2}, the $| \textbf{E} |^2$ profiles have been normalized for each geometry so that relative comparisons between the local and nonlocal results can be made. Because of this, it is not possible to directly compare the results for different geometries, but such comparisons will be considered below. In all cases, the $| \textbf{E} |^2$ values are qualitatively similar, both inside and around the structures. Quantitatively, however, the nonlocal enhancements are clearly lower. This is especially true for the triangular nanowires.

In order to accurately assess the $| \textbf{E} |^2$ enhancements around the nanowires, the precise maximum and average values were determined; Table \ref{Tb:nanowire_enhancements}.
\begin{table}[H]
  \caption{Maximum and average $| \textbf{E} |^2$ enhancements at the LSPR energies around cylindrical, square, and triangular nanowires with diameters or side-lengths of 50 nm. Each was illuminated with the $\textbf{E}$-field polarized along the longest axis of the structure. Distances from the nanowire surfaces over which averages were obtained are specified.}
  \label{Tb:nanowire_enhancements}
  \begin{ruledtabular}
    \begin{tabular}{c c c c c}
      \textbf{Nanowire Shape} & Max.\ & Avg.\ at 0.5 nm & Avg.\ at 1.0 nm & Avg.\ at 2.0 nm \\
      \hline
      Cylindrical (local) & 8.64 & 2.42 & 2.47 & 2.40 \\
      Cylindrical (nonlocal) & 7.85 & 2.32 & 2.39 & 2.34 \\
      Square (local) & 60.58 & 3.54 & 3.33 & 3.01 \\
      Square (nonlocal) & 39.79 & 3.02 & 2.91 & 2.69 \\
      Triangular (local) & 145.77 & 5.49 & 4.90 & 4.18 \\
      Triangular (nonlocal) & 71.40 & 3.42 & 3.30 & 3.01 
    \end{tabular}
  \end{ruledtabular}
\end{table}
%
The average values refer to fields averaged over certain distances from the nanowire surfaces. In all cases, decreases in both quantities are seen in the nonlocal results (as could also be inferred from Fig.\ \ref{fig:nanowires_E2}, and the discussion above). However, for the cylindrical nanowires, these decreases are negligible. It is also interesting to note that the average enhancements are higher 1.0 nm away from the surface than they are at 0.5 nm. For the square nanowires, the decreases are noticeably larger. There is approximately a 10\% difference in both quantities at 1.0 nm, and a 6\% difference in the average values at 2.0 nm. The decrease in the difference between average enhancements at a further distance from the surface is expected, as the near-fields that contribute to this exponentially decay. The decreases in $| \textbf{E} |^2$ enhancements for the triangular nanowires are strikingly larger than for the other geometries. For example, decreases of 104\% and 61\% in the maximum and average values, respectively, are seen at 0.5 nm. 

Considering that some physical processes are dependent on $| \textbf{E} |^4$ enhancements \cite{VanDuyne_LSPR-Overview_2007}, such as surface-enhanced Raman scattering (SERS), the differences between local and nonlocal electrodynamics could have significant implications for the interpretations of results. This statement is based on the fact that the nonlocal calculations are, in principle, more rigorous than the local ones. For example, if the actual electromagnetic contribution to SERS is smaller than expected on the basis of local theory, it is possible that chemical effects play a more important role than has been considered in the past \cite{SERS-rev_Nie_2008}. Such results are also likely to play a large role in the accurate interpretation of electron energy loss measurements for anisotropic nanoparticle structures, which have recently received attention within the framework of local electrodynamics \cite{SP-energy-dist_Schatz_2009}.

\section{Summary and Outlook}
\label{Section:Summary}

In summary, we detailed our electrodynamics method to calculate the optical response of an arbitrarily shaped structure described by a spatially nonlocal dielectric function. The formulation was based on converting the hydrodynamic Drude model into an equation of motion for the conduction electons, which then served as a current field in the Maxwell--Amp\`{e}re law. By discretizing this equation using standard finite-difference techniques, we incorporated it into a self-consistent computational scheme along with the standard equations used in the finite-difference time-domain method. 

Using the example of a cylindrical Au nanowire studied in our previous Letter \cite{McMahon_NLDiel}, we demonstrated the remarkable accuracy of our method through comparisons to analytical results. As new applications, we calculated the optical responses of thin metal Au films, Au nanowires, and spherical Au nanoparticles. These calculations demonstrated a number of effects that result from the spatial nonlocality in the dielectric response, including anomalous absorption, blueshifting of localized surface plasmon resonances, and decreases in electromagnetic field enhancements.

The results presented demonstrate the importance of including nonlocal effects when describing metal--light interactions at the nanometer length scale. It is presently difficult to compare our results directly with existing experimental studies, because most of these have involved heterogeneous collections of particles or non-continuous systems, which in the small size limit tend to average over nonlocal effects. It is our hope that these results will motivate new, and more precise experimental studies, particularly those on isolated nanostructures, where nonlocal effects are likely to play a large role.

In the future, we  plan to derive more accurate expressions than the hydrodynamic Drude model. By incorporating exchange and correlation effects, we also plan to compare nonlocal calculations directly with quantum mechanical approaches, such as electronic structure theory. Such expressions will allow more even more accurate descriptions of nonlocal optical phenomena than were presented in this work.

\appendix*
\section{Nonlocal Finite-Difference Equations}
\label{ap:nl_fd_eqs}

In this appendix, the finite-difference equations used to model nonlocal dielectric effects are derived, and their implementation is discussed.

First, the temporal derivatives in Eqs.\ (\ref{eq:FaradayLawrt}) and (\ref{eq:HD2L_AmpereLaw_rt}) are discretized using a leapfrog algorithm \cite{Taflove_FDTD},
\begin{equation}
  \label{eq:FaradayLawrt_leapfrog}
  \mu_0 \frac{\Hfieldr^{n+1/2} - \Hfieldr^{n-1/2}}{\Delta t} = - \curlF{\Efieldr^n}
\end{equation}
\begin{equation}
  \label{eq:HD2L_AmpereLaw_leapfrog}
  \varepsilon_0 \varepsilon_{\infty} \frac{\Efieldr^{n+1} - \Efieldr^n}{\Delta t} + \sum_j \JLjvecr^{n+1/2} + \JHDvecr^{n+1/2} = \curlF{\Hfieldr^{n+1/2}}
\end{equation}
where the superscript $n$ denotes a discrete time step. Equations (\ref{eq:current_Lorentz_TD-ADE}) and (\ref{eq:current_HD_TD-ADE}) are discretized using central finite-differences (necessary because of the second-order derivatives) centered at time-step $n$,
\begin{multline}
  \label{eq:current_Lorentz_discretized_TD-ADE}
  \frac{\JLjvecr^{n+1} - 2 \JLjvecr^n + \JLjvecr^{n-1}}{\Delta t^2} + 2 \delta_{\text{L}j} \frac{\JLjvecr^{n+1} - \JLjvecr^{n-1}}{2 \Delta t} + \omega_{\text{L}j}^2 \JLjvecr^n = \\
  \varepsilon_{0} \Delta \varepsilon_{\text{L}j} \omega_{\text{L}j}^{2} \frac{\Efieldr^{n+1} - \Efieldr^{n-1}}{2 \Delta t}
\end{multline}
\begin{multline}
  \label{eq:current_HD_discretized_TD-ADE}
  \frac{\JHDvecr^{n+1} - 2 \JHDvecr^n + \JHDvecr^{n-1}}{\Delta t^2} + \gamma \frac{\JHDvecr^{n+1} - \JHDvecr^{n-1}}{2 \Delta t} - \beta^{2} \nabla^2 \JHDvecr^n = \\
  \varepsilon_{0} \omega_{\text{D}}^{2} \frac{\Efieldr^{n+1} - \Efieldr^{n-1}}{2 \Delta t} ~~ .
\end{multline}

Next, update equations for $\JLjvecr$ and $\JHDvecr$ are obtained by rearranging Eqs.\ (\ref{eq:current_Lorentz_discretized_TD-ADE}) and (\ref{eq:current_HD_discretized_TD-ADE}),
\begin{equation}
  \label{eq:current_Lorentz_update-eq}
  \JLjvecr^{n+1} = \alpha_{\text{L}j} \JLjvecr^n + \xi_{\text{L}j} \JLjvecr^{n-1} + \eta_{\text{L}j} \frac{\Efieldr^{n+1} - \Efieldr^{n-1}}{2 \Delta t}
\end{equation}
where
\begin{equation}
  \label{eq:alpha-coeff_Lj}
  \alpha_{\text{L}j} = \frac{ 2 - \omega_{\text{L}j}^{2} \Delta t^2 }{ 1 + \delta_{\text{L}j} \Delta t }
\end{equation}
\begin{equation}
  \label{eq:xi-coeff_Lj}
  \xi_{\text{L}j} = -\frac{ 1 - \delta_{\text{L}j} \Delta t }{ 1 + \delta_{\text{L}j} \Delta t }
\end{equation}
\begin{equation}
  \label{eq:eta-coeff_Lj}
  \eta_{\text{L}j} = \frac{ \varepsilon_{0} \Delta \varepsilon_{\text{L}j} \omega_{\text{L}j}^{2} \Delta t^2 }{ 1 + \delta_{\text{L}j} \Delta t } ~~ ,
\end{equation}
and
\begin{equation}
  \label{eq:current_HDrude_update-eq}
  \JHDvecr^{n+1} = \alpha_{\text{HD}} \JHDvecr^n + \xi_{\text{HD}} \JHDvecr^{n-1} + \eta_{\text{HD}} \frac{\Efieldr^{n+1} - \Efieldr^{n-1}}{2 \Delta t}
\end{equation}
where
\begin{equation}
  \label{eq:alpha-coeff_HD}
  \alpha_\text{HD} = \frac{ 4 + 2 \Delta t^2 \beta^{2} \nabla^2  }{ 2 + \gamma \Delta t }
\end{equation}
\begin{equation}
  \label{eq:xi-coeff_HD}
  \xi_\text{HD} = -\frac{ 2 - \gamma \Delta t }{ 2 + \gamma \Delta t }
\end{equation}
\begin{equation}
  \label{eq:eta-coeff_HD}
  \eta_\text{HD} = \frac{ 2 \varepsilon_{0} \omega_\text{D}^2 \Delta t^2 }{ 2 + \gamma \Delta t } ~~ .
\end{equation}
%
%
%
Note that $\alpha_\text{HD}$ is an operator, rather than a simple coefficient. To use Eqs.\ (\ref{eq:current_Lorentz_update-eq}) and (\ref{eq:current_HDrude_update-eq}) in Eq.\ (\ref{eq:HD2L_AmpereLaw_leapfrog}), $\JLjvecr$ and $\JHDvecr$ are centered at time-step $n + 1/2$ by averaging,
\begin{equation}
  \label{eq:current_Lorentz_center-eq}
  \JLjvecr^{n+1/2} = \frac{ \JLjvecr^{n+1} + \JLjvecr^{n} }{2}
\end{equation}
\begin{equation}
  \label{eq:current_HDrude_center-eq}
  \JHDvecr^{n+1/2} = \frac{ \JHDvecr^{n+1} + \JHDvecr^{n} }{2} ~~ .
\end{equation}

Equations (\ref{eq:HD2L_AmpereLaw_leapfrog}), (\ref{eq:current_Lorentz_update-eq}), and (\ref{eq:current_HDrude_update-eq}) all contain $\Efieldr^{n+1}$. To obtain a consistent update, Eqs.\ (\ref{eq:current_Lorentz_center-eq}) and (\ref{eq:current_HDrude_center-eq}) [using Eqs.\ (\ref{eq:current_Lorentz_update-eq}) and (\ref{eq:current_HDrude_update-eq})] are inserted into Eq.\ (\ref{eq:HD2L_AmpereLaw_leapfrog}) and rearranged,
\begin{equation}
  \label{eq:E_update-eq_disc}
  \Efieldr^{n+1} = \frac{1}{\zeta_1 + \zeta_2} \bigg[ \zeta_1 \Efieldr^{n} + \zeta_2 \Efieldr^{n-1} + \curlF{\Hfieldr^{n+1/2}} - \textbf{J}_{\text{T}} (\xvec)^{n,n-1} \bigg]
\end{equation}
where
\begin{equation}
  \label{eq:E_update-eq_disc_coeff1}
  \zeta_1 = \frac{\varepsilon_0 \varepsilon_{\infty}}{\Delta t}
\end{equation}
\begin{equation}
  \label{eq:E_update-eq_disc_coeff2}
  \zeta_2 = \frac{1}{4 \Delta t} \left( \sum_j \eta_{\text{L}j} + \eta_\text{HD} \right)
\end{equation}
and
\begin{multline}
  \label{eq:J_for_E-update_disc}
  \textbf{J}_{\text{T}} (\xvec)^{n,n-1} = \frac{1}{2} \bigg\{ \sum_j \bigg[ \left( \alpha_{\text{L}j} + 1 \right) \JLjvecr^{n} + \xi_{\text{L}j} \JLjvecr^{n-1} \bigg] + \\
  \left( \alpha_\text{HD} + 1 \right) \JHDvecr^{n} + \xi_\text{HD} \JHDvecr^{n-1} \bigg\} ~~ .
\end{multline}

Rearrangement of Eq.\ (\ref{eq:FaradayLawrt_leapfrog}) gives the appropriate update equation for \Hfieldr,
\begin{equation}
  \label{eq:FaradayLawrt_leapfrog_disc}
  \Hfieldr^{n+1/2} = \Hfieldr^{n-1/2} - \frac{ \Delta t }{\mu_0} \curlF{\Efieldr^n} ~~ .
\end{equation}

In order to satisfy Eqs.\ (\ref{eq:Gaussrt}) and (\ref{eq:GaussMagrt}), a Yee spatial discretization \cite{Yee_FDTD} is used for the components of $\Efieldr$ and $\Hfieldr$ (i.e., they are offset and circulate one another). The $\JLjvecr$ and $\JHDvecr$ components are centered at the same spatial locations as the corresponding $\Efieldr$ components. All of the spatial derivatives in Eqs.\ (\ref{eq:current_HDrude_update-eq}), (\ref{eq:E_update-eq_disc}), (\ref{eq:J_for_E-update_disc}), (\ref{eq:FaradayLawrt_leapfrog_disc}) and are approximated using central finite-differences.

In order to model an arbitrarily shaped structure, the $\JLjvecr$ and $\JHDvecr$ components only exist at the grid positions of the corresponding nonlocal material. By not updating the currents outside of the structure, the additional boundary condition of Pekar is imposed \cite{Fuchs_ABC-review} -- i.e., the total nonlocal polarization current vanishes outside of the structure.

Equations (\ref{eq:current_Lorentz_update-eq}), (\ref{eq:current_HDrude_update-eq}), (\ref{eq:E_update-eq_disc}), and (\ref{eq:FaradayLawrt_leapfrog_disc}) form the complete and consistent set necessary to solve Eqs.\ (\ref{eq:MaxwellAmperert}) -- (\ref{eq:GaussMagrt}) for materials described by the constitutive relationship in Eq.\ (\ref{eq:E-to-D-k}) with the dielectric function given in Eqs.\ (\ref{eq:electric_permittivity}) -- (\ref{eq:hydrodynamic-Drude}).

\begin{acknowledgments}
J.M.M. and G.C.S. were supported by AFOSR/DARPA Project BAA07-61 (FA9550-08-1-0221), and the NSF MRSEC (DMR-0520513) at the Materials Research Center of Northwestern University. Use of the Center for Nanoscale Materials was supported by the \text{U. S.} Department of Energy, Office of Science, Office of Basic Energy Sciences, under Contract No.\ DE-AC02-06CH11357. 
\end{acknowledgments}


\end{document}